\documentclass[11pt,notitlepage]{article}
\title{\bf {Personnal recollections about the
birth of String Theory}\thanks{LPTENS-08/61}}

\author{\bf {Eugene Cremmer} \\
 Laboratoire de Physique Th\'eorique
de l'Ecole Normale Sup\'erieure\thanks{UMR8549
ENS/CNRS associ\'ee \`a l'Universit\'e Pierre et
Marie Curie} \\
 24 rue Lhomond, 75231 PARIS Cedex05,
France}

\date{November 2008}

\begin{document}

\maketitle

\vskip 3cm

\begin{center}

Invited contribution to the collective book:

\medskip

{\bf {The Birth of String Theory}}

\medskip

edited by 

Andrea Cappelli, Elena Castellani, Filipo
Colomo and Paolo Di Vecchia

\end{center}

\vfill\eject

In 1969, I was finishing my Ph.D. in Orsay under
the supervision of Michel Gourdin on
phenomenological works related to $e^{+} e^{-}$
annhilations. These works allowed me to deepen my
knowledge in particle physics as well as to learn
mastering difficult calculations. However, I was
more attracted by more formal research (that we
would like being able to say more fundamental!). 
The theoretical physics laboratory in Orsay was
also hit by the explosion of activity which
followed the paper of Veneziano, and a group of
people began to work on dual models: Bouchiat,
Gervais, Nuyts, Amati who was spending a sabbatical in
Orsay, as well as younger researcher Neveu, Scherk
and Sourlas. My first encounter with dual models
was to rebel against this too fashionable growing 
activity.
With Jean Nuyts I asked if there could be some
other example of a s-t dual amplitude with only
poles in both channels: we found an example with
poles lying on a logarithmic trajectory instead of
a linear one~\cite{C1}. However this amplitude led to many
unsatisfactory physical conclusions and I joined
the main stream.

\bigskip

\section{Factorizing the Pomeron with Jo\"el
Scherk at CERN}

In 1970, much work was devoted to the dual
multiloop amplitude, in particular by Kaku and
Yu~\cite{KY}.
Lovelace~\cite{L1} and Alessandrini~\cite{A1} showed 
the relation
between these amplitudes and the Neumann function
associated with a sphere with handles. I then
began a rather technical work, the study of a
multiloop operator and its dual
properties~\cite{C2}. I had
the chance to get a fellow position at the theory
division at CERN for 2
years (1971 and 1972). This was the first chance
of my career. The second one was to "integrate" the
dual group at CERN which was led by Daniele Amati.
In this group, there was not only a very
stimulating scientific atmosphere, but also a very
friendly one.
At that time the theory division had more than
a hundred people and the fellows who were not 
associated with a
group could easily stay isolated. Then arrived my
third and most important chance, the venue of
Jo\"el Scherk and the beginning of our
collaboration. I knew of course Jo\"el from Orsay
and had had discussions with him although not
belonging to the same "team" (Bouchiat-Meyer for
Jo\"el, Gourdin for me), but I had had no opportunity
to work with him. As early as that stage, Jo\"el was
a profound physicist. He was a very quiet and
efficient physicist. His notes were very clear and
are still readable many years later (contrary to
mine!), the detailed calculations  were always
accompanied by some commentary.

\bigskip

After some works on multicurrents dual
amplitude~\cite{CS1,CS2},
where we had some mathematical fun with domain
variational theory on Riemann surfaces, we came
back to a less technically demanding problem but 
more irritating one. The one-loop diagrams (and
multiloop) are constructed from unitarity and
should have analytic properties resulting from
unitarity (namely unitarity cuts and eventually
poles). However in the non-planar orientable, it
appears in the channel with zero isospin
(associated with the "pomeron") a new singularity
violating unitarity. Lovelace~\cite{L2} showed that 
when the
intercept is 1  this singularity is factorisable,
furthermore he conjectured that in 26 dimensions 2
sets of oscillators are cancelled by gauge
conditions leading to a modification of the
amplitude which has no longer cut but poles so
that unitarity is no longer violated. This
conjecture will be shown later.
 When the intercept is
one (implying the existence of a massless vector
particle) and the dimension is less or equal to 26 
(for pure
bosonic model), the dual resonance model is
ghost free (the time dimension is eliminated by the
gauge invariance associated with the intercept 1).
In 26 dimensions, null states appear that should
also be eliminated. The complete formula for the
loop amplitude was to be proven later after the projector
on the physical states had been constructed.
Starting with the conjectured form of the
amplitude~\cite{CS3}
we factorized the reggeon and the
pomeron poles simultaneously. The pomeron sector 
looks quite
similar to the Shapiro-Virasoro model with a slope
half that of the reggeon trajectory and an intercept
equal to 2 (implying a massless spin 2 particle). 
In the string language, this will correspond to
the transition from an open string to a closed string.
From the 2-non planar loops diagram with 3
external reggeons, it is possible to extract a 
3-pomeron vertex. The appearance of this sector of
new particles shows that the Veneziano model (open
strings) is not consistent alone and that we
must also include the Shapiro-Virasoro model
(closed strings) 
This has been done independently by Clavelli and 
Shapiro~\cite{CL} who have extended it to the
Neveu-Schwarz-Ramond model in 10 dimensions.

\bigskip

\section{Combining and splitting Strings with
Jean-Loup Gervais in Orsay}

Back in Orsay in 1973, I began a 
collaboration with Jean-Loup Gervais. The
introduction of the string picture had improved
tremendously our physical understanding of dual
models. Associated with the string picture is the
functional approach to dual theories. The initial
works of Hsue, Sakita and Virasoro~\cite{HBV} and those of
Gervais and Sakita ~\cite{GS1,GS2} were plagued by 
their inability
to project out the ghost states but explicitly
exhibited two important properties of dual models
which are not so transparent in the operator
formalism, namely duality and the connection
between loop amplitudes and Neumann functions.
As far as the Veneziano model is concerned, a
crucial progress was made by Goddard et
al.~\cite{GGRT} who
showed that the Lagrangian of the free
relativistic string being gauge-invariant, one
needs only to quantize the transverse components
of the string variable if one chooses the
appropriate gauge. Gervais and Sakita~\cite{GS3} 
subsequently
wrote the path integral associated with transition
probabilities of strings with this gauge
condition, in such a way that one can perform the
functional integration and obtain the original
Veneziano amplitude. Later Mandelstam~\cite{MA}, starting
from this amplitude, gave a complete prescription
for dealing with external excited states as
described by Goddard et al.. He proved that the
resulting amplitude is Lorentz-invariant only at
26 space-time dimensions and that the
three-reggeon vertex coincides with the one given
by Ademollo et al (ADDF)~\cite{ADDF}, thus establishing complete
connection between the string formalism and the
operator formalism of the Veneziano model. Kaku
and Kikkawa~\cite{KK}  have introduced a
multi-string formalism in a consistent way 
such that the topological structure of the
corresponding perturbation series is identical
with the structure of the dual theory, each dual
amplitude being obtained as a sum of several
Feynman graph contributions. This formalism, based
on a functional treatment of the string variable,
remained ambiguous because of the lack of precise
definition of the functional integration while a
careful determination of the functional
integration measure is necessary to obtain
Lorentz-invariant amplitudes that coincide with
dual amplitudes. Gervais and I overcome this
problem by developing an infinite component field
theory of interacting relativistic strings starting
from operator approach. In particular we
introduced a 3-string vertex either for  2 incoming
strings $\rightarrow $ 1 outgoing string (combining strings) or
for 1 incoming string $\rightarrow $ 2 outgoing
strings
(splitting strings). It was simply defined as the
overlapping of the 3 strings at a given time. We
showed that it was related to the ADDF three-reggeon
vertex  by allowing each of the 3 strings to propagate 
for a very long time~\cite{CG1}. We then
showed~\cite{CG2} that in
order to get the correct 2 incoming strings
$\rightarrow $ 2
outgoing strings amplitude, it was necessary to
add a direct 4-string interactions to the sum of 2 Feynman diagrams constructed
from the 3-string vertex and a propagator. This
defined 
the
4-string vertex introduced by Kaku and Kikkawa.
who had shown that three- and four-strings
vertices were sufficient in the tree
approximation. The formalism to all orders can be
defined but it is necessary to also introduce  an
infinite component field associated to the closed
string and in addition to closed strings
vertices also define a transition vertex between open and
closed strings. Although satisfactory from the point
of view of precise definition, this formalism is
very hard to use in practice as was already
seen in the 4-string amplitude.

\bigskip

\section{Compactifying Strings with Jo\"el Scherk
at LPTENS}

In october 1974, a group of physicists of the
theoretical physics laboratory in Orsay
(essentially the Bouchiat-Meyer group that I
joined on my return from CERN) moved to Paris and
founded the "Laboratoire de Physique Th\'eorique
de l'Ecole Normale Sup\'erieure".

A very elegant feature of dual models is to
predict the dimension of space-time namely 26 for
the Veneziano and Shapiro-Virasoro models and 10
for the Neveu-Schwarz-Ramond model. Unfortunately
these predictions are rather unphysical, moreover
these models are predicting zero mass particles
and are therefore incompatible with hadronic
physics. It is worth to remember that the same
"avatar" happened to Yang-Mills theory. This led
Scherk and Schwarz~\cite{SS1} as well as
Yoneya~\cite{YO} to study the
connection between dual models and general
relativity in particular in the zero slope limit
which was known to make connection between dual
models and field theory. 
In 1975, Scherk and Schwarz~\cite{SS2} made the
really daring proposal that dual models should be
interpreted as a quantum theory of gravity unified
with the other forces between quarks and
gluons. They suggested that considering some of
the dimensions to be compact does not lead to any
contradiction within the framework of dual models.
Scherk and I proved that this assertion was indeed
correct~\cite{CS4}. This was the beginning of a new 
and very
fruitful collaboration with Jo\"el. We defined the
theory of open and closed strings on a compact
space (chosen to be a hypertorus). In a 
 field theory on a compact space the momenta in
 the compactified directions (hyper torus of radii 
 $R_i$) is quantized ( $p_i = {n_i}/{R_i}$) and
 with
 a single field is associated an infinite
 Kaluza-Klein multitower of fields in lower
 dimension. For open strings this is also the only
 change. For closed strings the change is less
 trivial, although simple. One must introduce 
 another integer number (winding number)  $m_i$ 
 corresponding to how
 many times a string wraps around the torus before
 closing. The new states corresponding to the
 quantum numbers $n_i$ and $m_i$ get now an
 additional mass $M_i$ given by

 $$ M_i^2 = {n_i^2}/{R_i^2} +
 {m_i^2}{R_i^2}/{\alpha '^2}
 $$

 We showed that the corresponding modification
 to the computation of loops like replacing the
 integration on the momentum flowing in the loop by
 a summation on the quantized momenta did not
 affect all the good results like the absence of
 non physical singularities in the non planar
 orientable loop and the appearance of new
 particles associated to the compactified closed
 string. I must confess that our field theory
 prejudice on the limit R infinite and R nul
 prevented us to discover the T-duality of
 closed string theory, namely the complete symmetry

 $$ n_i \to m_i , m_i \to n_i , R_i \to {\alpha
 '}/{R_i} $$

 Generalized to the torus associated with a group,
 this compactification of closed strings was to
 lead to  the construction of the heterotic
 string~\cite{HET} in 1985. It is important also to note 
 that this
 kind of compactification is equivalent to
 introduction of quantum numbers in string theory
 by Bardacki and Halpern~\cite{BH}.

\bigskip

At that time, I turned to  a new direction of work 
still with Jo\"el Scherk - until his unfortunate
death in 1980 - and others. 
This became my Supergravity Era but 

{\centerline {\bf This is another story}}

\vskip 1cm

\noindent {\bf \large Acknowledgements}

\bigskip

It is a pleasure to thank the organizers/editors
to invite me to contribute to this volume. It is
also a pleasure to thank Nicole Ribet for a careful
reading of the manuscript as well as for useful
comments.

\end{document}